\documentclass[secnumarabic,aps,prl,reprint,groupedaddress]{revtex4-1}

\usepackage[english]{babel}
\usepackage{graphicx}
\usepackage{color}
\usepackage{amsmath}
\usepackage{tabularx}

\begin{document}
	
\newcommand{\unit}[1]{\:\mathrm{#1}}            
\newcommand{\To}{\mathrm{T_0}}
\newcommand{\Tp}{\mathrm{T_+}}
\newcommand{\Tm}{\mathrm{T_-}}
\newcommand{\EST}{E_{\mathrm{ST}}}
\newcommand{\Rp}{\mathrm{R_{+}}}
\newcommand{\Rm}{\mathrm{R_{-}}}
\newcommand{\Rpp}{\mathrm{R_{++}}}
\newcommand{\Rmm}{\mathrm{R_{--}}}
\newcommand{\ddensity}[2]{\rho_{#1\,#2,#1\,#2}} 
\newcommand{\ket}[1]{\left| #1 \right>} 
\newcommand{\bra}[1]{\left< #1 \right|} 

\title{Non-classical spin transfer effects in an antiferromagnet}
\author{Alexander Mitrofanov}
\author{Sergei Urazhdin}
\affiliation{Department of Physics, Emory University, Atlanta, GA, USA.}

\begin{abstract}
We simulate scattering of electrons by a chain of antiferromagnetically coupled quantum Heisenberg spins, to analyze spin-transfer effects not described by the classical models of magnetism. Our simulations demonstrate efficient excitation of dynamical states that would be forbidden by the semiclassical symmetries, such as generation of multiple magnetic excitation quanta by a single electron. Furthermore, quantum interference of spin wavefunctions enables generation of magnetization dynamics with amplitudes exceeding the transferred magnetic moment. The efficiency of excitation is almost independent of the electron spin polarization, and is governed mainly by the transfer of energy. Non-classical spin transfer may thus enable efficient electronic control of antiferromagnets not limited by the classical constraints.

\end{abstract}

\maketitle

{\bf Introduction.} Spin transfer (ST) effect - the transfer of spin from the itinerant electrons to magnetic systems - has provided unprecedented insights into nanomagnetism, and enabled efficient magnetoelectronic nanodevices~\cite{Ralph20081190,Kent2015}. The speed of ST-based devices utilizing ferromagnets (Fs) is limited by their characteristic dynamical frequencies, while the efficiency - by the requirement that the transferred spin is comparable to the spin of the nanomagnet~\cite{Ralph20081190}. These limitations can be alleviated in nanodevices based on antiferromagnets (AFs), thanks to the vanishing bulk magnetization and the high characteristic dynamical frequencies that are typically two orders of magnitude larger than in Fs~\cite{doi:10.1063/1.4918990, RevModPhys.90.015005, Moriyama2018, Gomonay2018,PhysRevLett.113.196602, Jungwirth2016,elezn2018,Gomonay2018,Nmec2018}. Furthermore, AFs are immune to perturbations by magnetic fields. These features may enable nanoscale $THz$ oscillators~\cite{PhysRevLett.116.207603, Khymyn2017,Kampfrath2010}, fast and resilient AF-based memory devices~\cite{PhysRevB.91.064423,PhysRevLett.99.046602,Moriyama2018}, and high-speed AF domain wall motion~\cite{Gomonay2018,PhysRevB.101.144431}.

While there are many similarities between ST effects in Fs and AFs, substantial differences are also expected. The magnetization of Fs can be well approximated as a semi-classical vector field. In particular, in the ground state of a Heisenberg F, the local spins are aligned~\cite{White2007}. For a simple collinear AF, the equivalent would be the N\'eel state, where the spins of two magnetic sublattices are aligned in the opposite directions. However, this state is not an eigenstate of the AF Heisenberg Hamiltonian~\cite{bethe1931theorie}. Instead, its ground state can be described as a N\'eel state dressed with a large population of sublattice magnons - spin flips spread out on one of the magnetic sublattices~\cite{Kamra_2019}. Since the effects of ST on the dynamical magnetization states can be viewed as stimulated emission of magnons that occurs at a rate proportional to magnon populations~\cite{Berger1996,PhysRevB.69.134430,PhysRevLett.119.257201}, the magnon-dressed states of AFs may be affected by ST very differently from the pure N\'eel state.

The dressed N\'eel states of AFs originate from the non-commutativity of different components of spin~\cite{bethe1931theorie}. The non-commutativity of the spin components of the conduction electrons and of the rotations of AF spins have been introduced in the semiclassical approximation for AF order ~\cite{PhysRevB.86.245118, PhysRevB.98.134450}, but the contributions to ST resulting from the non-commutativity of AF spin components, which cannot be described semi-classically, remain unexplored. These contributions may be similar to the "quantum ST'' demonstrated for Fs, which also originates from the non-commutativity of different spin components~\cite{PhysRevB.69.134430,PhysRevLett.119.257201}. \emph{Here and below, we use the terms "quantum ST'', or equivalently "non-classical ST'', to refer to the contributions to ST that cannot be described within semiclassical approximation for magnetism, but instead require that the localized spins forming the magnetization are described by the Schrödinger equation.} Since the effects of spin non-commutativity are generally much larger in AFs than in Fs, more significant non-classical ST effects may be also expected. Indeed, the results described below show that non-classical ST effects may be  dominant in AFs, enabling efficient current-induced excitation of dynamical states even in the absence of semiclassical ST.

{\bf Model.} To analyze the quantum problem of interaction between the spin-polarized current and AF, we consider an itinerant electron initially propagating in a non-magnetic medium, and scattered by a 1D chain of local AF-coupled Heisenberg spins-1/2. The system can be described by the tight-binding Hamiltonian~\cite{PhysRevB.99.094431,petrovi2020spintronics,alex2020energy,supplemental}

\begin{equation}\label{eq:hamiltonian}
	\begin{split}
		\hat{H} = -\sum_{i}b|i\rangle\langle i+1|\\
		-\sum_{j}(J_{sd}|j\rangle\langle j|\otimes \hat{\mathbf{S}}_j\cdot\hat{\mathbf{s}}
		-J\hat{\mathbf{S}}_j\cdot\hat{\mathbf{S}}_{j+1}),
	\end{split}
\end{equation}
where $i$ enumerates the tight-binding sites of the entire system, $j$ - the sites occupied by the localized spins-1/2 representing the AF, $\hat{\mathbf{s}}$, $\hat{\mathbf{S}}_j$ are the spin operators of the electron and the local spins. The first term in Eq.~(\ref{eq:hamiltonian}) describes hopping of the itinerant electron, the second - exchange interaction between the itinerant electron and the local spins, and the last term - exchange interaction between localized spins. Periodic boundary conditions for both the electron and the spin chain are used to avoid reflections at the boundaries.

The evolution of the system is determined by numerically integrating the time-dependent Schrödinger equation with the Hamiltonian Eq.~(\ref{eq:hamiltonian})~\cite{supplemental}. The observable quantities are determined using the density matrices $\hat{\rho}_e=Tr_m\hat{\rho}$ and $\hat{\rho}_m=Tr_e\hat{\rho}$ for the itinerant electron and the local spins, obtained by tracing out the full density matrix $\hat{\rho}$ with respect to the other subsystem~\cite{PhysRevB.99.094431}. The expectation value of observable $\hat{A}$ pertaining to the electron is $\left<\hat{A}\right>=Tr(\hat{A}\hat{\rho}_e)$, while the probability of its value $a$ is $P_a=\left<\psi_a|\hat{\rho}_e|\psi_a\right>$, where $\psi_a$ is the corresponding eigenstate. The quantities pertaining to AF are determined similarly.

\begin{figure}
	\includegraphics[width=\columnwidth]{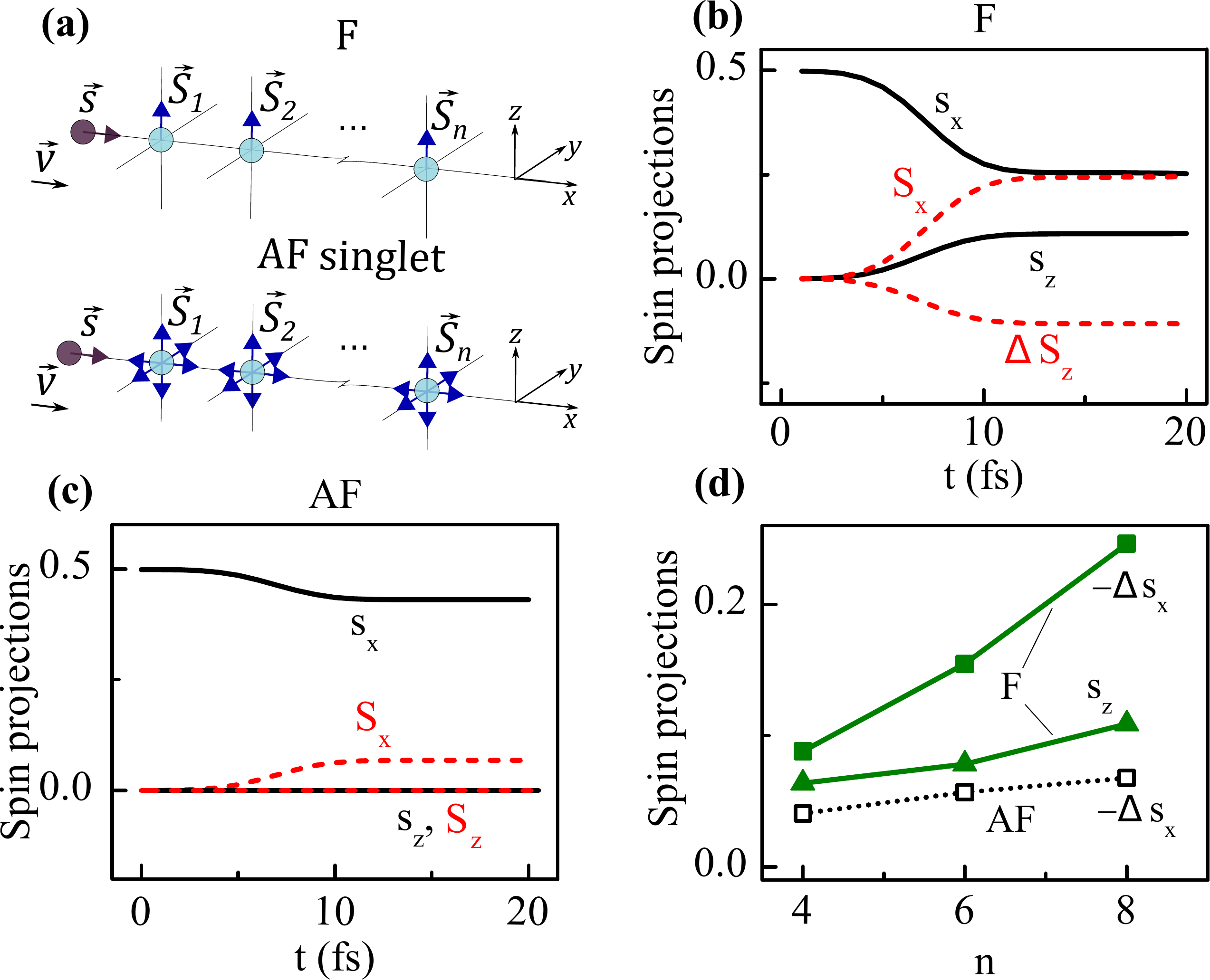}	
	\caption{\label{fig:F_AF} (Color online) (a) Schematics of the simulated systems that consist of an itinerant electron scattered by F (top) and AF (bottom), initially in their ground states. (b,c) Evolution of the expectation values of $x$- and $z$-components of the electron spin $\mathbf{s}$ and the total spin $\mathbf{S}$ of the magnetic system that consists of $n=8$ spins-1/2, for F (b), and AF (c). $\Delta S_z$ is the variation relative to $S_z=4$. (d) Dependence of the transferred spin on the number $n$ of local spins, for F and AF, as labeled. The simulations were performed using $b=1$~eV, $J_{sd}=0.1$~eV, and $J=-0.1$~eV ($0.1$~eV) for F(AF), with the scattered electron initially forming a Gaussian wave packet centered at the wavenumber $k_0=5nm^{-1}$.}
\end{figure}

{\bf Spin transfer in AF vs F.} First, we compare the ST effects in AF to those in F modeled using Eq.~(\ref{eq:hamiltonian}) with the opposite sign of $J$. Both systems are initialized in their ground states - F spins aligned with the z-axis, and AF spins forming a spin singlet~\cite{bethe1931theorie, Karbach_1998}. We note that all the components of the local spins vanish in the spin singlet state, so it cannot be described semiclassically. Thus, ST in this state is purely quantum.
 
The electron is initialized, at time $t=0$, as a wave packet with spin along the x-axis, propagating in the non-magnetic medium towards the magnetic system [Fig.~\ref{fig:F_AF}(a)]. The spins of the electron and of the magnetic system start to vary at $t>5$~fs, Figs.~\ref{fig:F_AF}(b,c). The variations become negligible at $t>12$~fs, after the wave packet is completely scattered~\cite{supplemental}. The well-defined transitions among these regimes allow us to unambiguously quantify the ST effects.

In the simulations for F, the x-component of the electron spin orthogonal to the local spins becomes reduced, while the corresponding component for the local spins increases by the same amount [Fig.~\ref{fig:F_AF}(b)], consistent with the theories of ST~\cite{Slonczewski1996,Berger1996,Ralph20081190}. The electron spin also acquires a component along the z-axis, while the corresponding local spin component becomes reduced by the same amount, due to the quantum ST~\cite{PhysRevLett.119.257201,PhysRevB.99.094431,alex2020energy}.

In case of AF, the electron's initial spin is also partially transferred to the local spins [Fig.~\ref{fig:F_AF}(c)]. In contrast to F, the z-component of electron spin does not vary, consistent with the isotropic spin properties of the singlet state. The transferred spin increases with increasing size of the magnetic system for both F and AF, due to the increasing interaction time with the itinerant electron [Fig.~\ref{fig:F_AF}(d)]. The spin transferred to AF always remains smaller than the spin transferred to F.

\begin{figure}
	\includegraphics[width=\columnwidth]{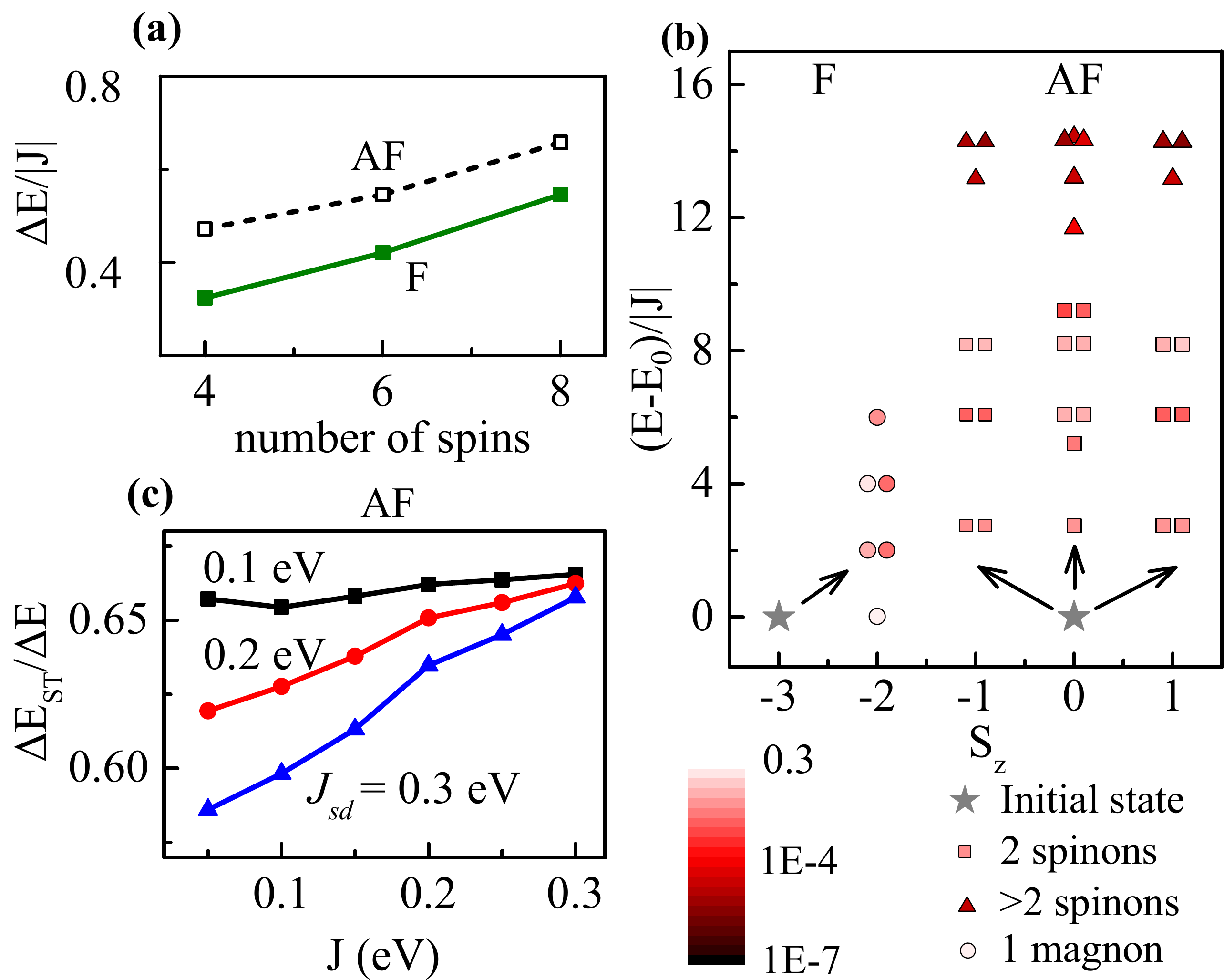}	
	\caption{\label{fig:excitations} (Color online) (a) Transfer of energy from electron to the local spins vs the chain length $n$. (b) Relative energies of the eigenstates for F (left) and AF (right) with $n=6$ vs their spin projection on the z-axis. Only the states with finite amplitudes after scattering are shown. Stars: ground state, circles: 1-magnon states (for F), squares:  2-spinon states (for AF), triangles: states with more than two spinons. Color scale: the probability of the state after scattering. Some symbols are slightly shifted for clarity. (c) The ratio of energy transferred to magnetic excitations with finite spin to the total transferred energy vs $J$, at the labeled values of $J_{sd}$. The simulation parameters and the initial states are the same as in Fig.~\ref{fig:F_AF}, unless specified otherwise. }
\end{figure}

{\bf Energy transfer and the spectrum of excitations.} ST is likely not the only effect controlling the current-induced dynamical processes in AFs. Indeed, the spin angular momentum in the ground state, such as the  N\'eel state of 3d AFs, is zero. In a gedanken experiment, N\'eel order can be reversed by exchanging the neighboring opposite spins via a superposition of two opposite N\'eel states. The net spin remains zero throughout this process, and thus does not require ST. However, switching between stable magnetic configurations requires that the system overcomes the energy barrier between them. Thus, energy transferred to the magnetic system must play an important role in magnetic reversal~\cite{alex2020energy}. 

The energy transferred from the scattered electron to the magnetic system is larger for AF than F [Fig.~\ref{fig:excitations}(a)]. We reconcile this result with the weaker ST in AF by analyzing the dynamical magnetization states induced by the electron scattering. We use Bethe ansatz to classify the eigenstates of the magnetic systems in terms of the elementary excitations - magnons for F, and spinons - fractionalized spin-1/2 quasiparticles - for the 1D AF~\cite{bethe1931theorie, Karbach_1998,supplemental}. The final state of the magnetic system is projected onto these eigenstates to determine the probabilities of their excitation.

The energies of the eignestates with non-zero amplitudes after electron scattering are plotted in Fig.~\ref{fig:excitations}(b) versus the z-component of their spin, for $n=6$. For $F$, all $6$ of the eigenstates excited by ST are 1-magnon states, with $S_z=-2$. This is expected from angular momentum conservation, since each magnon carries spin 1, so a spin-1/2 electron can excite at most one magnon.

In contrast to F, a variety of multi-quasiparticle eigenstates are excited in AF. For an integer-spin chain, spinons must be generated in pairs, with the possible $z$ spin component $-1$, $0$ or $1$. All these possibilities are realized in the studied system [Fig.~\ref{fig:excitations}(b)]. In contrast to F, spin conservation does not limit the number of generated quasiparticles, as long as their spins add up to $0$ or $1$. Indeed, $11$ of the $31$ eigenstates of AF excited by ST contain more than 2 spinons [triangles in Fig.~\ref{fig:excitations}(b)]. These results are consistent with many-spinon excitation observed in neutron scattering~\cite{Dalla_Piazza2015, Mourigal2013}.

The results of Fig.~\ref{fig:excitations}(b) explain why energy transfer in AF can be more efficient than in F, even though ST is less efficient. In F, spin conservation limits the accessible dynamical magnetization states, and since each magnon carries the same spin $1$, magnon excitation is directly tied to ST. For AFs, excitation of many different dynamical states is allowed by spin conservation. They can have different spin directions, adding up to smaller net spin transfer. The relative significance of ST can be characterized by the ratio $\Delta E_{ST}/\Delta E$ of energy transferred to the states with $S_z=\pm 1$ to the total transferred energy~\cite{comment}. The value of $\Delta E_{ST}/\Delta E$ varies with the system parameters such as exchange interaction [Fig.~\ref{fig:excitations}(b)]. While the variations observed in these simulations are modest, a large range of efficiency of non-ST excitation may be likely achieved by varying the spectrum of magnetic excitations.

\begin{figure}
	\includegraphics[width=\columnwidth]{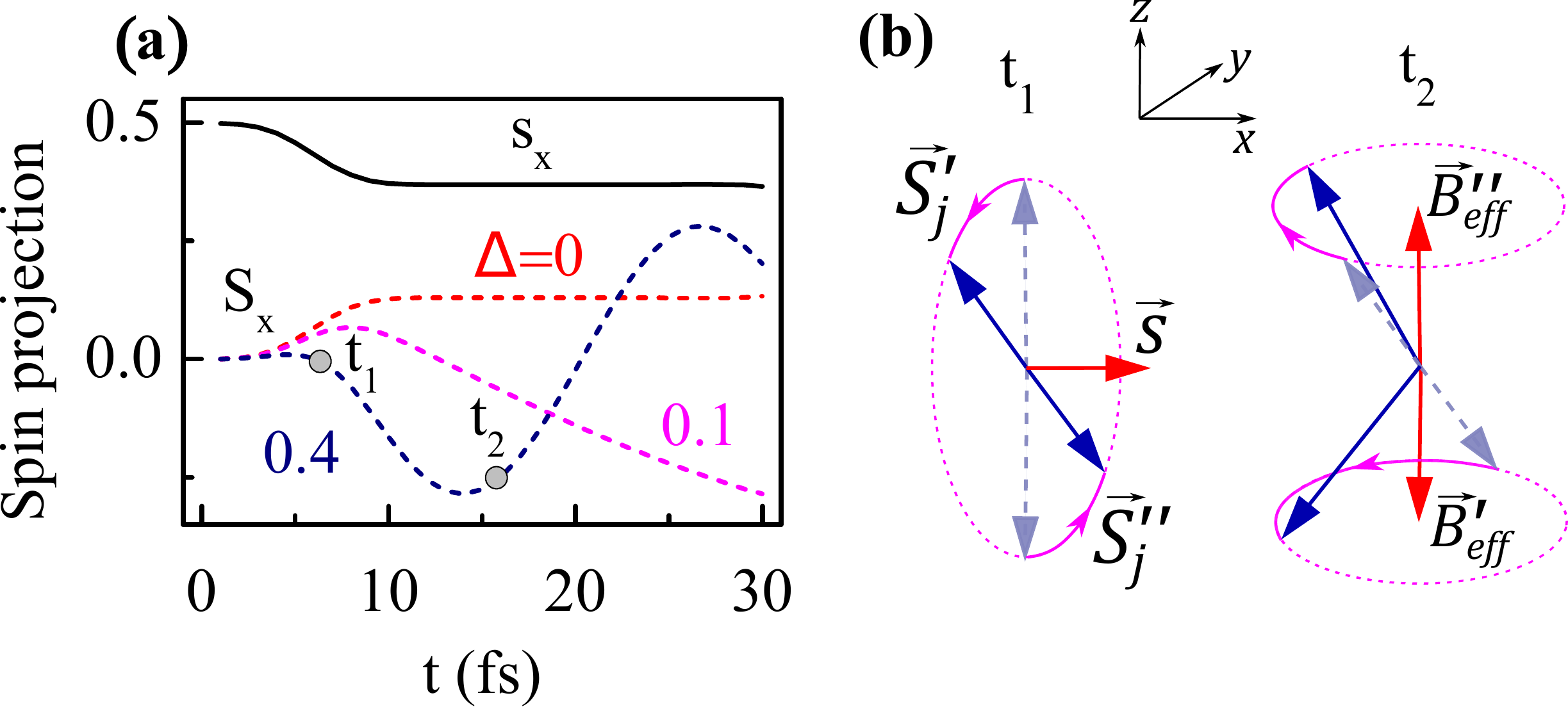}	
	\caption{\label{fig:2_Neel} (Color online) ST for an anisotropic AF chain with $n=4$, initially in the superposition of two Néel states. (a) Evolution of the expectation values of the $x$ spin components of electron (solid curve) and the local spins (dashed curves), at the labeled values of anisotropy $\Delta$. (b) Schematics of spin-up (labeled $\mathbf{S}'_j$) and spin-down (labeled $\mathbf{S}''_j$) components of the wavefunction for one of local spins at $\Delta=0.4$, at times labeled $t_1$ and $t_2$ in panel (a).	Curved arrows and dashed circles show the trajectories of the corresponding spin wavefunction components during ST (left) and after ST (right). $B'_{eff}$ and $B''_{eff}$ are the effective anisotropy fields experienced by $\mathbf{S}'_j$ and  $\mathbf{S}''_j$, respectively. For $\Delta=0.4$, ST is compensated by the anisotropy torques at $t_1$. These effects are not shown for clarity.}
\end{figure}

{\bf Enhancement of dynamical amplitude due to spin interference.} The possibility to generate magnetic dynamics without ST may enable current-induced excitations with much larger amplitudes, and consequently a higher efficiency of current-induced magnetic switching, than would be achievable with only ST-mediated excitations. This possibility is demonstrated in Fig.~\ref{fig:2_Neel} for an anisotropic AF chain of four spins initialized in the state $(\ket{\uparrow\downarrow\uparrow\downarrow}-\ket{\downarrow\uparrow\downarrow\uparrow})/\sqrt{2}$, which is a superposition of two N\'eel states, an excited eigenstate with two spinons~\cite{supplemental}.

The XXZ-type spin-anisotropy of the chain is introduced by adding the term $J\Delta\sum_{j}\hat{S}_j^z\hat{S}_{j+1}^z$ to the Hamiltonian Eq.~(\ref{eq:hamiltonian}). Figure~\ref{fig:2_Neel}(a) shows the spin evolution for different values of $\Delta$. By symmetry, the sum of the y- and z-components of both local spin wavefunction components remain zero. The  dependence $s_x(t)$ is nearly identical for all three shown values of $\Delta$ [solid curve in Fig.~\ref{fig:2_Neel}(a)]. In contrast, the evolution of $S_x$ is strongly dependent on $\Delta$. For $\Delta=0$, it mirrors the evolution of the electron's spin, as expected for the isotropic Hamiltonian. For $\Delta=0.1$ and $0.4$, $S_x$ first slightly increases, and then starts to oscillate with amplitude significantly larger than the transferred spin. The period of the oscillation is larger for $\Delta=0.1$, so the oscillation appears as a monotonic variation in Fig.~\ref{fig:2_Neel}(a). 

The mechanism enabling large-amplitude dynamics driven by small ST is illustrated in Fig.~\ref{fig:2_Neel}(b) for one of the local spins. Exchange torque exerted by the itinerant electron's spin $\vec{s}$ results in the rotation of the spin-up component $\mathbf{S}'_j$ of the local spin wavefunction away from the y-axis, while the spin-down component  $\mathbf{S}''_j$ rotates towards the y-axis (top schematic). There is no ST associated with these opposite rotations, because the effects of these rotations cancel each other. Thus, large rotation angles can be achieved without violating angular momentum conservation. 

The rotated spin components precess around the effective anisotropy fields. For  $\mathbf{S}'_j$, the effective field $B'_{eff}$ is directed mostly down, resulting in its clockwise precession around the z-axis, Meanwhile, for $\mathbf{S}''_j$, the corresponding field $B''_{eff}$ is up, resulting in the opposite sense of precession. Consequently,  the interference between $\mathbf{S}'_j$ and $\mathbf{S}''_j$ periodically varies between constructive [Fig.~\ref{fig:2_Neel}(b)] and destructive, resulting in the oscillation of $S_x$ with amplitude exceeding ST.

This amplitude enhancement mechanism requires that i) the initial state contains a superposition of two reversed N\'eel states, and ii) the spin precession sense of these states is opposite. Such enhancement may thus be expected for 3D AFs with uniaxial anisotropy, since their ground state contains a large weight of the reversed N\'eel state~\cite{Kamra_2019}, and the uniaxial anisotropy field reverses upon reversal of AF spins.

\begin{figure}
	\includegraphics[width=\columnwidth]{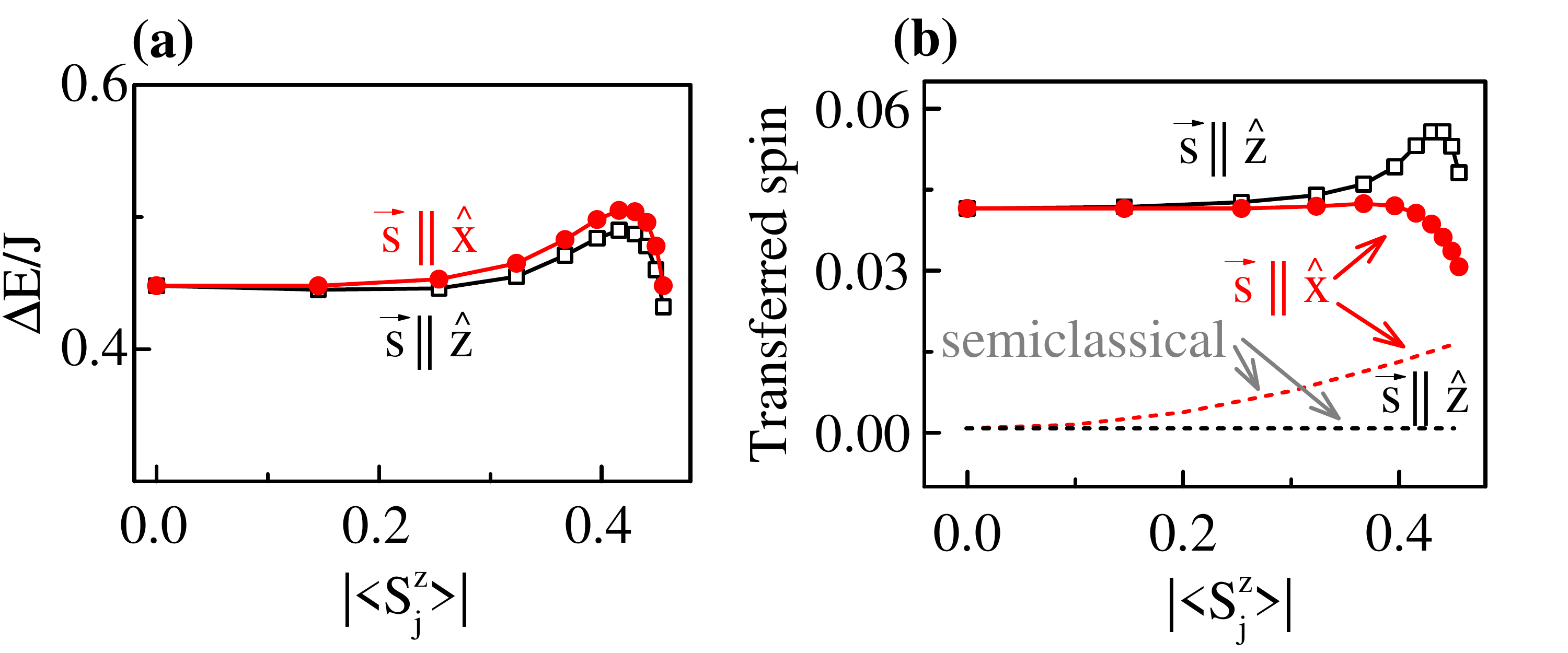}	
	\caption{\label{fig:staggered_field} (Color online) Dependence of the energy transfer (a) and ST (b) on the degree of N\'eel ordering imposed by the staggered Zeeman field and on the polarization of the incident electron, as labeled. The ordering is characterized by the magnitude $|\left<S^z_j\right>|$ of the z-component of one of the local spins. Dashed lines in (b): ST calculated using the adiabatic semiclassical approximation for the local spins~\cite{supplemental}.}
\end{figure}

{\bf Effects of magnetic ordering and relation to semiclassical ST.} The effects described above were obtained for the purely quantum states of 1D AF characterized by the vanishing expectation values of all the components of local spins. To show that these effects are relevant to the N\'eel states, we add a staggered Zeeman term $\sum_{j}(-1)^j\gamma\hat{S}_j^zB_{st}$ to the Hamiltonian Eq.~(\ref{eq:hamiltonian}). As the staggered field $B_{st}$ is increased, the magnitudes $|\left<S^z_j\right>|$ of the local spin z-components increase, at large $B_{st}$ approaching the semiclassical N\'eel limit~\cite{supplemental}. 

Energy transfer exhibits a modest dependence on the degree of N\'eel ordering, and is almost independent of the electron polarization relative to the N\'eel vector [Fig.~\ref{fig:staggered_field}(a)], likely due to the accessibility of many excited states even in the N\'eel limit. ST is larger when the incident electron spin is collinear with the N\'eel vector, $s_z||\hat{z}$, when the classical contribution to ST is absent [Fig.~\ref{fig:staggered_field}(b)]. However, even for  $s_z\perp\hat{z}$ maximizing the classical ST, the quantum contribution is dominant except for $|\left<S^z_j\right>|\rightarrow 0.5$, when it becomes reduced due to the increasing gap in the excitation spectrum neglected in the semiclassical approach~\cite{supplemental,note2}.

{\bf Conclusions.} We utilized simulations of electron scattering by a quantum AF spin chain, to elucidate ST effects not described by the semiclassical approximation for the magnetization. Our main result is the dominance of excitation processes that are not accounted for by ST alone, and involve dynamical states that cannot be described semiclassically. For instance, a variety of many-quasiparticle eigenstates are excited by a single electron, thanks to the existence of magnetic modes with different spins that can add up to the total spin of $0$ or $1$ required by the angular momentum conservation. The excitation of spin-0 eigenstates, which accounted for almost half of the excited states in our simulations, is not associated with ST, and is governed instead by the energy transfer. We expect these insights to be relevant to 2D and 3D AFs, which also exhibit excitation modes with different spins, e.g. spin-up and spin-down magnons in uniaxial AFs.

We showed that the availability of non-classical dynamical states allows generation of magnetization dynamics with amplitudes exceeding the transferred magnetic magnetic moment, thanks to spin interference. Furthermore, our analysis of ST in the spin-singlet ground state of 1D AF, a specific case of a spin liquid - a correlated spin state that cannot be described semiclassically - charts a path for the analysis of ST in spin liquids and other non-classical spin states~\cite{Balents2010,Mourigal2013}, with possible applications in neuromorphic systems~\cite{Chen2020}.

Our simulations reveal that ST is nearly independent of the conduction electron's spin polarization, thanks to the availability of multiple spin transfer channels regardless of the electron's spin polarization. In contrast, in the semiclassical approximation, ST vanishes when the electron is polarized parallel to N\'eel vector. This warrants a re-examination of the prior analyses based on the expected symmetry of ST in AFs~\cite{PhysRevLett.123.247206, Bodnar2018, PhysRevLett.125.077201, Shi2020}.

The demonstrated effects may also contribute to phenomena related to spin transport, such as spin diffusion~\cite{Manchon2017SD}, giant magnetoresistance~\cite{PhysRevB.73.214426}, the spin Hall effects in AF systems~\cite{Manchon2017SMR,PhysRevB.97.014417, PhysRevLett.113.196602}. Our results also suggest that spin currents can be efficiently converted into spin excitations in AFs~\cite{Hirobe2017,lebrun2020longdistance,Han2020}, regardless of the electron spin polarization, in stark contrast to the highly anisotropic spin conversion in Fs~\cite{Handbook2016}. As a consequence, the spin diffusion length in metallic AFs should be almost independent of the electron spin polarization relative to the magnetic order. However, it should exhibit a strong dependence on the energy of the conduction electrons. 

Finally, we discuss the expected effects of finite temperature. The constraints imposed by energy transfer should be diminished for the low-lying states, but remain significant for magnetically ordered AFs, due the large gap in their excitation spectrum. The other demonstrated effects, associated with the availability of efficient semiclassically forbidden excitation channels, should be unaffected by finite temperatures. 

This work was supported by the U.S. Department of Energy, Office of Science, Basic Energy Sciences, under Award \# DE-SC0018976.

\bibliography{AFM_QST}

\begin{thebibliography}{48}
\bibitem [{com()}]{comment}%
  \BibitemOpen
  \href@noop {} {}\bibinfo {note} {To verify this, one can use x as the
  quantization axis for spinons, so that only the x spin component is involved
  in ST by symmetry. Eigenstate classification in terms of spinons remains the
  same.}\BibitemShut {Stop}%
\end{thebibliography}
\bibliographystyle{apsrev4-1}

\end{document}